\documentclass[twocolumn,superscriptaddress,amssymb,amsmath,nobibnotes, aps,prab,10pt,floatfix]{revtex4-1}
\expandafter\ifx\csname package@font\endcsname\relax\else
\expandafter\expandafter
\expandafter\usepackage
\expandafter\expandafter
\expandafter{\csname package@font\endcsname}%
\fi
\usepackage{hyperref}
\usepackage{graphicx}
\usepackage{booktabs}
\usepackage{amssymb}
\usepackage{amsfonts}
\usepackage{amsmath}
\usepackage{float}
\usepackage{textcomp}
\usepackage{epstopdf}
\usepackage{gensymb}
\usepackage{blindtext}
\usepackage{siunitx}
\usepackage{tabularx}
\usepackage{enumitem}
\usepackage{subfig}
\usepackage{fourier} 
\usepackage{makecell}
\usepackage{tikz}

\usepackage[margin=0.5in]{geometry}

\hypersetup{pdfnewwindow}

\begin{document}
\title{Mitigation of the Microbunching Instability Through Transverse Landau Damping}
\author{A.~D.~Brynes}\email[]{alexander.brynes@elettra.eu}
\affiliation{Elettra-Sincrotrone Trieste S.C.p.A., 34149 Basovizza, Trieste, Italy}
\author{G.~Perosa}\email[]{Now at FREIA Laboratory, Uppsala University, Uppsala, Sweden}
\affiliation{Elettra-Sincrotrone Trieste S.C.p.A., 34149 Basovizza, Trieste, Italy}
\affiliation{University of Trieste, Dept. Physics, 34127 Trieste, Italy}
\author{C.-Y.~Tsai}
\affiliation{Department of Electrotechnical Theory and Advanced Electromagnetic Technology, Huazhong University of Science and Technology, Wuhan, 430074, China}
\author{E.~Allaria}
\affiliation{Elettra-Sincrotrone Trieste S.C.p.A., 34149 Basovizza, Trieste, Italy}
\author{L.~Badano}
\affiliation{Elettra-Sincrotrone Trieste S.C.p.A., 34149 Basovizza, Trieste, Italy}
\author{G.~De~Ninno}
\affiliation{Elettra-Sincrotrone Trieste S.C.p.A., 34149 Basovizza, Trieste, Italy}
\affiliation{University of Nova Gorica, 5000 Nova Gorica, Slovenia}
\author{E.~Ferrari}
\affiliation{Deutsches Elektronen-Synchrotron, Notkestraße 85, 22607 Hamburg, Germany}
\author{D.~Garzella}
\affiliation{Elettra-Sincrotrone Trieste S.C.p.A., 34149 Basovizza, Trieste, Italy}
\author{L.~Giannessi}
\affiliation{Elettra-Sincrotrone Trieste S.C.p.A., 34149 Basovizza, Trieste, Italy}
\affiliation{Istituto Nazionale di Fisica Nucleare, Laboratori Nazionali di Frascati, 00044 Frascati, Rome, Italy}
\author{G.~Penco}
\affiliation{Elettra-Sincrotrone Trieste S.C.p.A., 34149 Basovizza, Trieste, Italy}
\author{P.~Rebernik~Ribi\v{c}}
\affiliation{Elettra-Sincrotrone Trieste S.C.p.A., 34149 Basovizza, Trieste, Italy}
\author{E.~Roussel}
\affiliation{Universit\'e Lille, CNRS, UMR 8523 - PhLAM - Physique des Lasers Atomes et Mol\'ecules, F-59000 Lille, France}
\author{S.~Spampinati}
\affiliation{Istituto Nazionale di Fisica Nucleare, Laboratori Nazionali di Frascati, 00044 Frascati, Rome, Italy}
\author{C.~Spezzani}
\affiliation{Elettra-Sincrotrone Trieste S.C.p.A., 34149 Basovizza, Trieste, Italy}
\author{M.~Trov\`o}
\affiliation{Elettra-Sincrotrone Trieste S.C.p.A., 34149 Basovizza, Trieste, Italy}
\author{M.~Veronese}
\affiliation{Elettra-Sincrotrone Trieste S.C.p.A., 34149 Basovizza, Trieste, Italy}
\author{S.~Di~Mitri}
\affiliation{Elettra-Sincrotrone Trieste S.C.p.A., 34149 Basovizza, Trieste, Italy}
\affiliation{University of Trieste, Dept. Physics, 34127 Trieste, Italy}

\begin{abstract} 
    The microbunching instability has been a long-standing issue for high-brightness free-electron lasers (FELs), and is a significant show-stopper to achieving full longitudinal coherence in the x-ray regime. This paper reports the first experimental demonstration of microbunching instability mitigation through transverse Landau damping, based on linear optics control in a dispersive region. Analytical predictions for the microbunching content are supported by numerical calculations of the instability gain and confirmed through the experimental characterization of the spectral brightness of the FERMI FEL under different transverse optics configurations of the transfer line between the linear accelerator and the FEL.
\end{abstract}

\maketitle

\section{Introduction}\label{sec:introduction}

Free-electron lasers (FELs), in particular those with a wavelength range at the extreme ultraviolet and below, require high-brightness electron bunches with a low slice energy spread, small transverse emittance, and a smooth current profile \cite{Schmuser, PhysRevSTAB.10.034801}. Meeting these constraints is necessary in order to provide high-quality photon pulses with a narrow bandwidth and a smooth spectrum, but these properties are difficult to achieve, in part due to the collective interactions that occur between charged particles contained within a small region of 6D phase space \cite{PhysRevSTAB.17.110702,PhysRep.539.1}. 

Once the electron beam is accelerated beyond an energy of approximately $100$\,\si{\mega\electronvolt}, the most significant drivers of phase-space dilution and beam non-uniformity are coherent synchrotron radiation (CSR) \cite{NIMA.398.2-3.373,NewJPhys.20.073035,TESLA-FEL-Report-1995-05}, longitudinal space charge (LSC) \cite{NIMA.393.1-3.376,PhysRevSTAB.11.034401} and geometric wakefields in accelerating structures \cite{IntJModPhysA.22.3736,PhysRevAccelBeams.22.014401}. In addition to driving transverse emittance growth \cite{NIMA.393.1-3.494,PhysRevE.51.1453}, in the case of the first of these effects, and causing a quadratic chirp in the longitudinal phase space \cite{PhysRevLett.112.044801}, in the case of the third, all of these effects can give rise to the microbunching instability \cite{PhysRevSTAB.18.030704,NIMA.528.1-2.355,NIMA.483.1-2.268,NIMA.528.1-2.355,PhysRevSTAB.11.034401,PhysRevSTAB.5.064401,PhysRevSTAB.5.074401,PhysRevSTAB.12.080704}. 

This instability arises due to non-uniformities in the electron bunch and shot noise in the low-energy injector of the accelerator \cite{PhysRevAccelBeams.20.054402}, which are then amplified by the collective effects mentioned above. All of these interactions between particles in the bunch are associated with a characteristic impedance \cite{CAS2018}. The impedance due to accelerating cavities is, in general, relatively small in the wavelength range of interest (below $100$\,\si{\micro\metre}) for a short-wavelength FEL, and therefore the LSC and CSR impedances are more significant contributors to the amplification of modulations in the beam. As a result of these impedances, driving an FEL using a bunch with a non-uniform longitudinal phase space can result both in a reduction in the photon pulse energy produced in the undulators and the spectral quality of the radiation, with microbunching-induced sidebands appearing in the spectrum \cite{PhysRevAccelBeams.19.050701,PhysRevAccelBeams.22.080702}.

Detailed calculations \cite{PhysRevSTAB.10.054403,PhysRevSTAB.10.104401} and simulation studies \cite{PhysRevSTAB.11.030701} have demonstrated that, in the case of linear accelerator (linac) based FELs, initial modulations with a wavelength in the range $\lambda_{0} \approx 10 \hbox{--} 100$\,\si{\micro\meter} undergo the largest amplification due to these collective effects. After the bunch is compressed longitudinally by a factor of approximately $10$, which is the standard operation mode of the FERMI FEL \cite{NatPhoton.6.699}, it is expected that the strongest modulations in the longitudinal profile of the bunch at the end of the linac should be around $1 \hbox{--} 10$\,\si{\micro\metre}, which is indeed in agreement with measurements \cite{SciRep.10.5059}. 

The most common method to mitigate the microbunching instability is through the use of a \lq laser heater\rq\,\cite{NIMA.528.1-2.355}, a technique which increases the energy spread of the bunch by impinging a laser pulse on the electron beam while travelling through an undulator magnet. Numerous VUV and x-ray FELs employ this technique regularly in order to optimize their performance \cite{PhysRevSTAB.13.020703,NIMA.843.39,PhysRevSTAB.17.120705,FEL2017.WEP018}; improvements in spectral quality and pulse intensity have been observed thanks to the use of laser heaters. 

This technique is sufficiently robust for improving the quality of FEL light produced that a number of facilities regularly use this method during machine operation. However, certain FEL schemes -- high-gain harmonic generation (HGHG) \cite{PhysRevA.44.5178}, for example -- require low beam energy spread, and so using the laser heater may not be the ideal solution to improve the FEL output in all cases. Furthermore, increasing the energy spread results in a reduction of the FEL gain and can be deleterious to the performance of a seeded FEL; additionally, installing a chicane, undulator and laser in the machine introduces extra costs to the facility which may be avoidable.

For this reason, a number of schemes have been proposed to suppress the microbunching instability based on control of the beam optics. Such schemes generally involve generating and exploiting transverse-to-longitudinal mixing, either through the use of dispersion generated by bending magnets, \cite{PhysRevLett.111.054801,PhysRevAccelBeams.23.014403}, transverse deflecting cavities \cite{PhysRevSTAB.15.022802}, or transverse gradient undulators \cite{NewJPhys.17.073028}; other examples include adjusting the bunch compression scheme \cite{NIMA.1050.168145}, or strongly focusing the beam, thereby enhancing the effect of intrabeam scattering \cite{NIMA.1048.167968,NewJPhys.22.083053,PhysPlasmas.28.013112}. As many of these schemes are based on symplectic transformations of the beam phase space, they are reversible, meaning that the transverse emittance and slice energy spread can, in principle, be recovered at the entrance to the FEL. Many of these schemes do not require additional hardware, and could therefore be attractive methods for future machines which are expected to be susceptible to the microbunching instability, in addition to providing another technique for existing machines. 

In spite of the large number of inventive proposals, phase mixing at high energy -- through control of the linear momentum compaction of a dispersive transfer line -- is to date the only demonstrated viable strategy to mitigate the instability \cite{PhysRevLett.112.134802}, also in the absence of a laser heater \cite{PhysRevAccelBeams.20.120701,PhysRevAccelBeams.23.110703}. This paper reports on the first demonstration of transverse Landau damping of the instability. The microbunching content is controlled by means of the horizontal beam properties and a suitable betatron motion in a high-energy dispersive transfer line. The technique is therefore complementary to that described above, and is capable of achieving a noticeable improvement of the FEL spectral brightness.  

In Sec.\,\ref{sec:theory}, we review the theoretical considerations necessary to understand the development of the microbunching instability. Sec.\,\ref{sec:setup} discusses the FERMI machine, and the methods employed to characterize the microbunching instability, including a new infrared spectrometer used to measure modulations in the longitudinal profile of the electron bunch across the wavelength range that is most pertinent for microbunching studies. Results of experimental measurements taken at FERMI for different transverse beam optics configurations are presented in Sec.\,\ref{sec:results}, using the spectrometer and the FEL performance for direct and indirect measurements of the impact of microbunching on the beam. In Sec.\,\ref{sec:discussion}, we discuss these results, demonstrating that changes in the transverse beam optics can cause a similar improvement in FEL performance to that which can be achieved with a laser heater; comparisons are made between the measurements and theoretical calculations, performed using a semi-analytic Vlasov solver to calculate the microbunching gain along the accelerator lattice. Finally, our conclusions are summarized in Sec.\,\ref{sec:conclusion}, and a future outlook for microbunching mitigation is explored. 

\section{Theory}\label{sec:theory}

The microbunching instability at a certain wavenumber $k$ and a longitudinal position along the machine $s$ is quantified using the bunching factor $b\left(k, s\right)$; in the case of a one-dimensional analysis of the current profile of the electron bunch, this quantity is given by the Fourier transform of the longitudinal bunch distribution. The gain in microbunching $G_{f}$ at this wavenumber is then calculated as the ratio between the final and initial bunching factors: $G_{f}\left(k, s_{f}\right) = \left|b_{f} \left(k, s_{f}\right) / b_{0} \left(k, 0\right)\right|$, where the subscript $0$ denotes the initial bunching factor and the subscript $f$ represents the final value \cite{NIMA.483.1-2.516,PhysRevSTAB.5.064401,PhysRevSTAB.5.074401}.

Given an initial bunching factor $b_{0}\left(k, 0\right)$ in the particle distribution, the microbunching instability develops as the bunch travels through the accelerator according to the following integral equation \cite{PhysRevSTAB.5.064401,PhysRevSTAB.5.074401}:

\begin{equation}\label{eq:bunching_factor}
    b\left[k(s), s\right] = b_{0}\left[k(s), s\right] + \int_{0}^{s} K(\tau, s) b\left[k(\tau), \tau\right] d\tau,
\end{equation}

\noindent where the kernel of the integral equation is given as follows:

\begin{equation}\label{eq:kernel}
\begin{split}
    K(\tau, s) = i k(s) R_{56}(\tau \to s) \frac{I(\tau)}{I_{A}}\frac{Z\left[k(\tau), \tau\right]}{\gamma_{0}} \exp\left[-\frac{k_{0}^2}{2} U(s, \tau)^{2} \sigma_{\delta 0}^{2}\right] \times \\
    \exp\left[-\frac{k_{0}^{2} \epsilon_{x0}^{G}}{2 \beta_{x0}} T(s, \tau)\right].
\end{split}
\end{equation}

\noindent In the above equation, the $R_{56}$ term gives the longitudinal dispersion in the $6\times 6$ linear transport matrix between two longitudinal positions $\tau$ and $s$; $I\left(\tau\right)$ is the current of the electron bunch at $\tau$; $\gamma_{0}$, $\sigma_{\delta 0}$, $\epsilon_{x 0}^{G}$, $\beta_{x 0}$ are the relativistic Lorentz factor, initial uncorrelated energy spread, geometric horizontal emittance and horizontal beta function of the beam, respectively; $Z\left[k\left(\tau\right), \tau\right]$ are the impedances due to collective effects; $k_{0}$ is the uncompressed wavenumber of the modulation; and $I_{A} \approx 17.045$\,\si{\kilo\ampere} is the Alfv\'en current. $k(s)$ is related to $k_{0}(s)$ via the compression factor $C(s)$. The optics functions $U(s, \tau)$ and $T(s, \tau)$ are calculated as:

\begin{subequations}
\begin{equation}\label{eq:u_r56}
    U\left(s, \tau\right) = C(s)R_{56}(s) - C(\tau)R_{56}(\tau),
\end{equation}
\begin{equation}
    T(s, \tau) = \left[\beta_{x 0} V(s, \tau) - \alpha_{x 0}W(s, \tau)\right]^2 + W(s, \tau)^2,
\end{equation}
\end{subequations}

\noindent with $\alpha_{x0}$ the initial horizontal alpha function and $V$ and $W$ given by:

\begin{subequations}
\begin{equation}
    V(s, \tau) = C(s)R_{51}(s) - C(\tau)R_{51}(\tau),
\end{equation}
\begin{equation}
    W(s, \tau) = C(s)R_{52}(s) - C(\tau)R_{52}(\tau).
\end{equation}
\end{subequations}

\noindent These expressions are the analogues to the relation for longitudinal dispersion in Eq.\,\ref{eq:u_r56} given above, calculated in terms of the horizontal dispersion and its derivative in the linear transport matrix. For our purposes, we do not need to consider the effects of vertical dispersion, although the expressions take the same form. 





It can be seen from Eq.\,\ref{eq:kernel} that there are two exponential damping terms which can be increased in order to reduce the amplification of modulations in the beam. As both of these terms are related to Landau damping -- the first and second as a result of the longitudinal and transverse properties of the beam and lattice, respectively -- we can label them $LD_{\parallel}$ and $LD_{\perp}$. The second of these can be recast as:

\begin{equation}\label{eq:landau_damping_transverse}
    LD_{\perp} = \exp\left[-\frac{k^{2}(s)\mathcal{H}_{x}(s)\epsilon_{x0}}{2\gamma(s)}\right],
\end{equation}

\noindent with the parameter $\mathcal{H}_{x}$ derived from the definitions of $T(s,\tau)$, $V(s,\tau)$ and $W(s,\tau)$:

\begin{equation}
    \mathcal{H}_{x}(s) = \frac{R_{51}^{2}(s) + \left[\beta_{x}(s) R_{52}(s) + \alpha_{x}(s) R_{51}(s)\right]^{2}}{\beta_{x} (s)}.
\end{equation}

\noindent This damping term arises due to the longitudinal mixing associated with the beam emittance and the dispersive elements of the transfer matrix. As can be seen from Eq.\,\ref{eq:kernel}, there are three terms (for constant values of accelerating gradient and beam transverse emittance) which can be increased in order to damp the growth of the bunching factor for a beam with initial bunching $k_{0}$ along the linac: the uncorrelated energy spread $\sigma_{\delta 0}$, the longitudinal dispersion $R_{56}$, or the H-function $\mathcal{H}_{x}$. 

As the bunching factor is inversely correlated with the slice energy spread of the electron bunch, the most common method to mitigate the microbunching instability is through the use of a \lq laser heater\rq\,\cite{NIMA.528.1-2.355}, a device consisting of a small four-dipole magnetic chicane with an undulator in its center. As the electron bunch passes through the undulator, a laser pulse is directed onto the bunch, thereby imparting a modulation in energy. When the beam passes through the second half of the chicane, the overlapping paths in longitudinal phase space traversed by the modulated electrons cause an increase in the energy spread, thereby damping the instability. Numerous VUV and x-ray FELs employ this technique regularly in order to optimize their performance \cite{PhysRevSTAB.13.020703,NIMA.843.39,PhysRevSTAB.17.120705,FEL2017.WEP018}; improvements in spectral quality and pulse intensity have been observed thanks to the use of laser heaters.

A recent experiment was conducted at FERMI \cite{PhysRevAccelBeams.23.110703} -- see Sec.\,\ref{subsec:machine} for some more details about the machine -- in which the $R_{56}$ of the linac-to-FEL transfer line (or \lq spreader\rq) was varied, and its effect on the FEL spectrum was observed. The previous nominal configuration for the spreader at FERMI was to have $R_{56} = 0$, the \lq isochronous\rq\, condition; this is based on the fact that the CSR-induced microbunching in the spreader is minimised for a smaller $R_{56}$ \cite{PhysRevAccelBeams.20.024401}. However, the competing effect that is driven by the isochronous optics is the lack of an additional term to damp the bunching factor (Eq.\,\ref{eq:kernel}), which can grow along the spreader due to the CSR impedance in dispersive regions and the ever-present LSC impedance. Given the improved FEL spectrum reported in Ref.\,\cite{PhysRevAccelBeams.23.110703} for a non-isochronous condition in the spreader, it can be argued that the phase mixing which arises due to the longitudinal slippage of particles that are modulated in density \cite{PhysRevLett.112.134802} has a more significant effect than CSR-induced microbunching under certain conditions. 

Another route towards mitigating the growth of the microbunching instability in dispersive regions using only the beam optics and existing hardware, as seen in Eq.\,\ref{eq:kernel}, is to increase the $\mathcal{H}_{x}$ function. 

\section{Experimental Setup}\label{sec:setup}

\subsection{Machine Configurations}\label{subsec:machine}

\begin{figure}[bth!]
	\begin{center}
		\includegraphics[width=8.6cm]{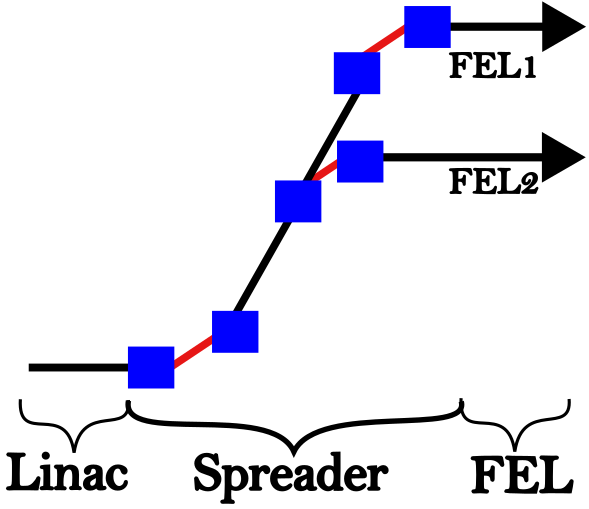}
		\caption{Schematic of the FERMI spreader. The electron beam exits the linac and then travels through the spreader line into either of the two FEL lines. Dipoles are shown in blue, and in between each double-bend achromat cell (the areas with red lines) the action of $LD_{\perp}$ can damp the microbunching instability. Other elements, such as quadrupole magnets, diagnostic devices and collimators, are not shown.} \label{fig:fermi_layout}
	\end{center}
\end{figure}

Experiments to study the impact of $\mathcal{H}_{x}$ on the microbunching instability were performed at the FERMI FEL facility in Trieste, Italy \cite{JSynchRad.22.485}. This machine consists of an RF photoinjector, linac (including a variable bunch compressor), transfer line (or \lq spreader\rq), and two FEL lines. The spreader line, as shown in Fig.\,\ref{fig:fermi_layout}, has two branches in order to thread the beam to the FEL sections, each consisting of two double-bend dipole cells separated by quadrupole magnets for beam optics control. 

Two sets of measurements were performed to determine the effect of the transverse beam optics in the spreader line, one on each of the FEL lines (see Table\,\ref{table:machine_setup} for a summary of the machine configurations). Although the final beam energy and compression factor was not the same for both measurements ($1240$\,\si{\mega\electronvolt} and $8$ in the case of FEL1, respectively, and $1535$\,\si{\mega\electronvolt} and $10$ on the FEL2 line), the transverse beam optics were similar along the linac, thanks to the use of optics matching routines \cite{PhysRevSTAB.15.012802,FEL2022.WEP04}. The beam Twiss parameters were then measured at the exit of the linac and after the spreader line, at the entrance to the undulator section.  

\begin{table}[bth!]
\centering
\caption{Machine setup for experiments on both FEL lines} 
\label{table:machine_setup}
\begin{tabular}{llll}
    \hline\hline
    \textbf{Parameter} & \textbf{FEL1} & \textbf{FEL2}  & \textbf{Unit} \\
    \hline\hline
    \multicolumn{4}{c}{\textsc{Linac}} \\
    \hline
    Bunch charge & \multicolumn{2}{c}{$500$} & \si{\pico\coulomb} \\
    Initial peak current & \multicolumn{2}{c}{$70$} & \si{\ampere} \\
    Initial energy spread & \multicolumn{2}{c}{$2$} & \si{\kilo\electronvolt} \\
    Compression factor & $8$ & $10$ & \hbox{--} \\
    Final peak current & $550$ & $700$ & \si{\ampere} \\
    Final beam energy & $1240$ & $1535$ & \si{\mega\electronvolt} \\
    Normalized transverse emittance & $1.2$ & $1.4$ & \si{\micro\metre\radian}\\
    \hline
    \multicolumn{4}{c}{\textsc{Spreader}} \\
    \hline
    Dipole bending angle & \multicolumn{2}{c}{$52$} & \si{\milli\radian} \\
    Dipole length & \multicolumn{2}{c}{$0.4$} & \si{\metre} \\
    \hline
    \multicolumn{4}{c}{\textsc{FEL}} \\
    \hline
    Seed laser wavelength & $270.5$ & $250$ & \si{\nano\metre} \\
    Harmonic & $12$ & $8\times 5$ & \hbox{--} \\
    FEL wavelength & $22.5$ & $6.25$ & \si{\nano\metre} \\
    \hline
\end{tabular}
\end{table}  

\begin{figure}
	\begin{center}
		\centering
		\subfloat[FEL1]{  
			\includegraphics[width=8.6cm]{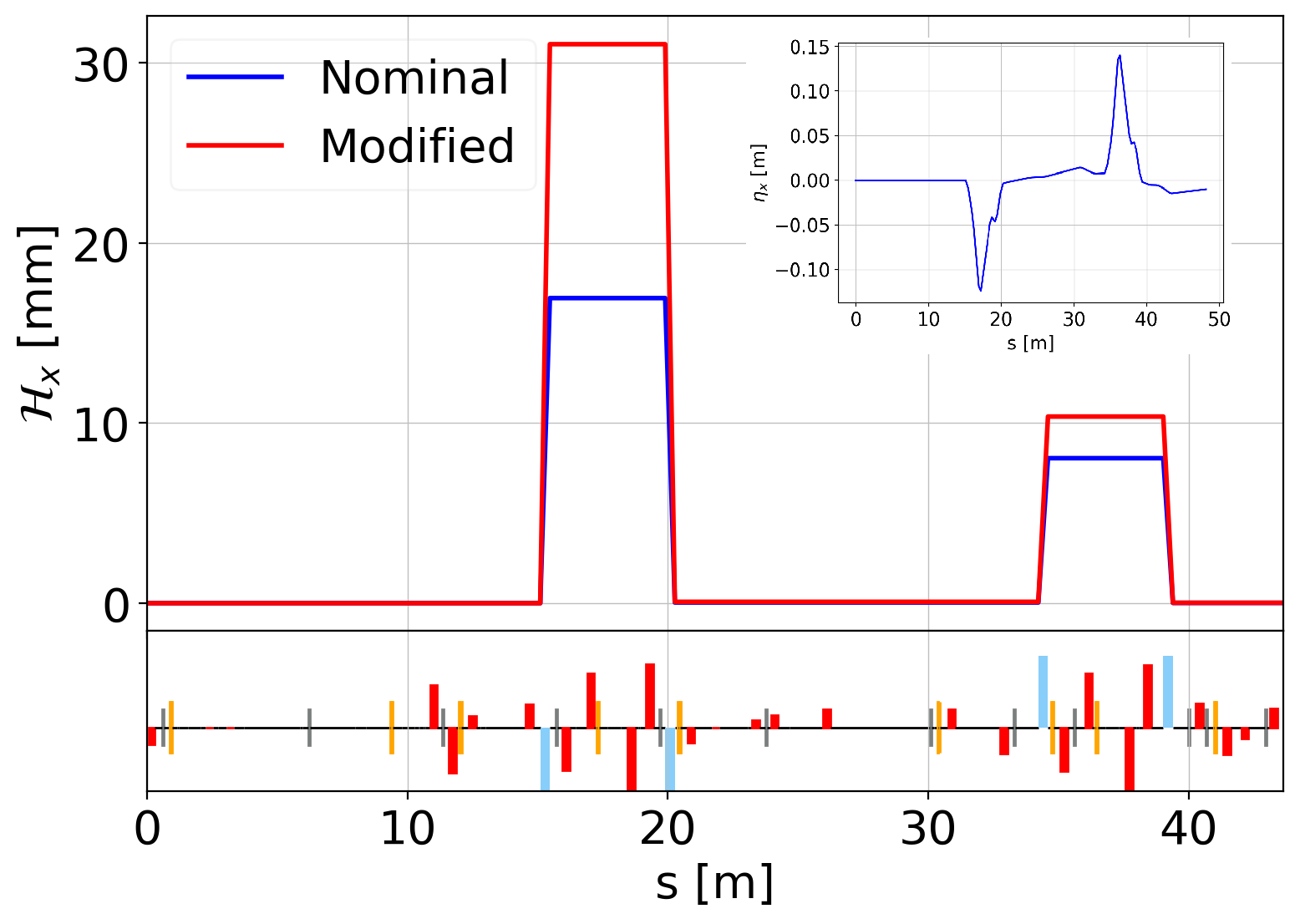} 
			\label{fig:hx_fel1}}
        \vfill
		\subfloat[FEL2]{  
			\includegraphics[width=8.6cm]{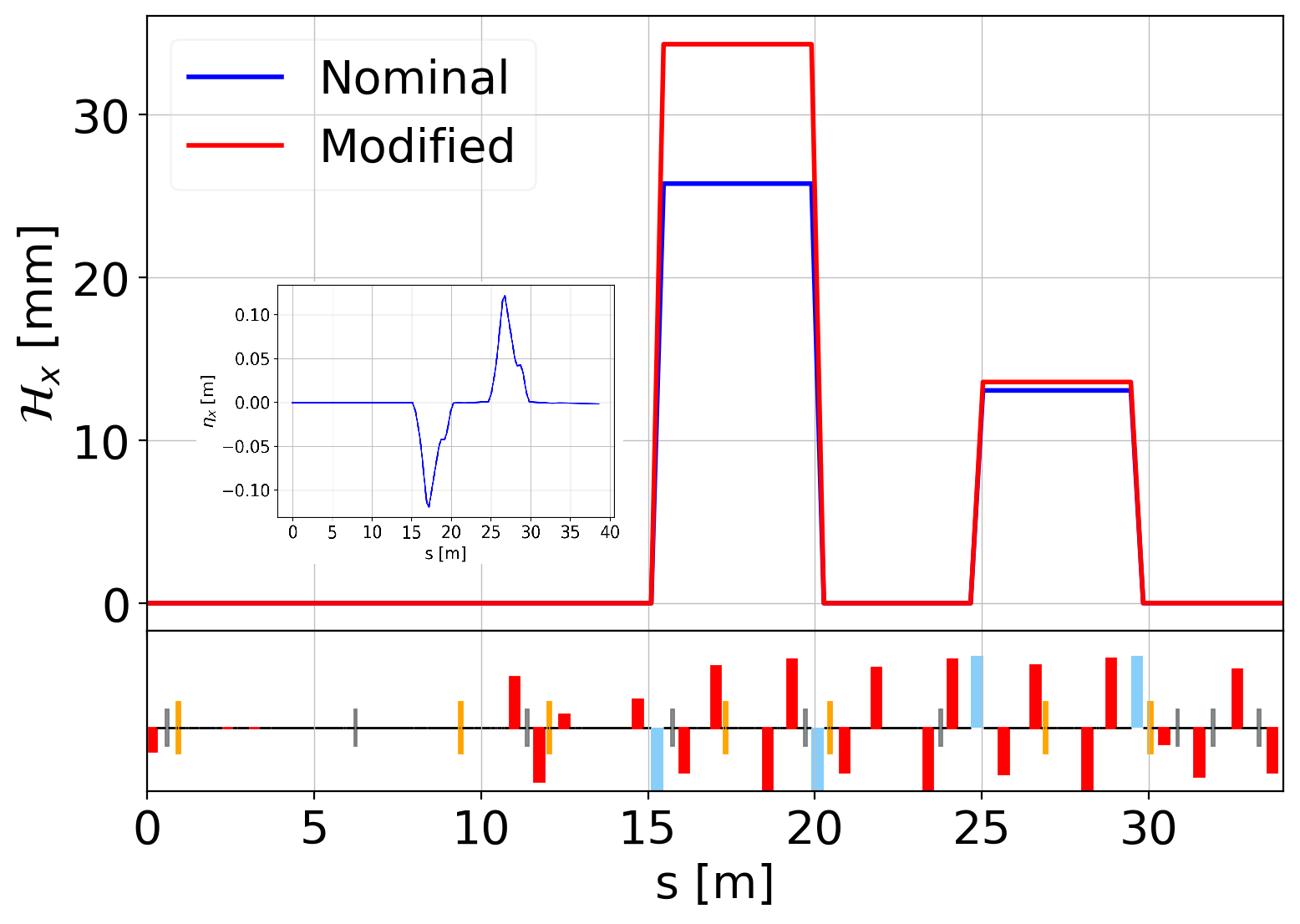} 
			\label{fig:hx_fel2}}
		\caption{$\mathcal{H}_{x}$ across both branches of the spreader line for two different transverse optics configurations. The total $R_{56}$ of the spreader for FEL1 and FEL2 was $1.23$\,\si{\milli\metre} and $0.78$\,\si{\milli\meter}, respectively. The dispersion function along each line is shown in the inset figures.} \label{fig:hx}
	\end{center}
\end{figure}

For each set of measurements on FEL1 and FEL2, two different transverse optics configurations were found. The nominal case is based on the principle of using optics balance to cancel CSR kicks in the two double-bend achromat cells in the spreader \cite{PhysRevLett.110.014801}; the second, modified condition achieved a larger value of $\mathcal{H}_{x}$ in the spreader by adjusting the strength of two quadrupole magnets before the entrance to the transfer line, with minimal change to the CSR cancellation condition. The values of $\mathcal{H}_{x}$ through each spreader line for the two sets of optics are plotted in Fig.\,\ref{fig:hx}. It can be seen that in the dispersive regions, the absolute maximum value of $\mathcal{H}_{x}$ with the modified optics is around twice as large in the case of FEL1, and $\approx 50$\,\si{\percent} larger for FEL2, when compared with the nominal settings. 

After transporting the beam through the spreader, the transverse beam parameters were then characterized in the subsequent matching section, located in front of the FEL modulators. In order to optimize the FEL performance, the electron beam optics through the FEL were modified using the intra-undulator quadrupole magnets to maximize the coupling between the beam and the radiation produced by the FEL interaction, and the FEL undulators were tapered in order to increase the intensity of the radiation pulse. The pulse intensity and spectral quality of the FEL was monitored by tuning the pulse energy, trajectory and timing of the laser used to seed the FEL interaction, and the strength of the dispersive magnet after the modulator undulator \cite{GiannessiFEL}, a procedure that is routinely followed at FERMI. The dispersive magnet strength and the seed properties were the same for both sets of beam optics. 

\subsection{Microbunching Measurement Methods}\label{subsec:methods}

\subsubsection{Indirect Methods}\label{subsubsec:indirect}

The FEL performance on both beamlines can be characterised through the use of PRESTO \cite{JSynchRad.23.35}, an on-line spectrometer which allows for measurement of the photon spectrum in real-time, and an intensity monitor based on the atomic photo-ionization of a rare gas when the photons pass through. Real-time data acquisition of electron beam and FEL diagnostics \cite{ICALEPCS17.THPHA044}, coupled with scans over machine parameters, are used to provide information about the best working point for the FEL. 

In addition to studying the total pulse energy, the spectrum quality is particularly useful in determining the microbunching content in the beam. As discussed in Ref.\,\cite{PhysRevAccelBeams.19.050701}, the microbunching instability can create sidebands in the spectrum, disturbing the smooth Gaussian pulse that is characteristic of a seeded FEL. Previous experiments undertaken at FERMI \cite{PhysRevAccelBeams.23.110703} demonstrated that a measurement of the ratio between the spectral intensity of the sidebands and the integrated intensity of the pulse could be used as an indirect measurement of the microbunching content in the beam. 

In the context of microbunching studies, one of the most useful actuators to scan is the laser heater pulse intensity, which can be controlled through the use of a polarizing attenuator. Numerous studies \cite{PhysRevSTAB.13.020703,PhysRevSTAB.17.120705,NIMA.843.39} have shown that an optimal working point for the FEL can be found if a small uncorrelated energy spread is added to the electron beam in the laser heater. For the experiments described above, after the FEL was optimized, the FEL pulse intensity and spectrum were measured as a function of the laser heater intensity. If no improvement is found in the FEL performance when the laser heater is used, then it can be inferred that there is a minimal amount of microbunching in the electron bunch. For this reason, we can look at the pulse intensity normalized to the maximum across a scan of the laser heater energy, in order to observe if the increase in energy spread imposed on the beam is damping the instability, and to disambiguate the results from other factors which may be impacting the FEL process. As a variation in the value of $\mathcal{H}_{x}$ has an analogous effect in terms of damping the instability to the action of the laser heater, it is expected that increasing $\mathcal{H}_{x}$ will require a smaller laser heater action to improve the FEL performance. 

\subsubsection{Direct Methods}\label{subsubsec:direct}

Using proxy methods to characterise the effect of microbunching on the electron beam, such as the FEL properties, is incredibly useful, since these properties are the most important for the users of the FEL pulses. Additionally, these methods are -- in general -- non-destructive, allowing simultaneous operation of the machine and FEL characterization without disturbing the electron beam. Nevertheless, direct measurement methods are necessary in order to give more detailed information about the small-scale structure within the bunch. 

Previous direct measurements of microbunched beams have utilised a transverse deflecting cavity, coupled with a bending magnet, to image the full longitudinal phase space of the electron bunch \cite{PhysRevSTAB.18.030704,SciRep.10.5059,PhysRevAccelBeams.23.104401}, from which the bunching factor as a function of wavelength can be measured using Fourier analysis. Such calculations, however, can involve detailed post-processing, and do not immediately yield information about the microbunching content in the beam. 

\begin{figure}[!ht]
	\begin{center}
		\includegraphics[width=86mm]{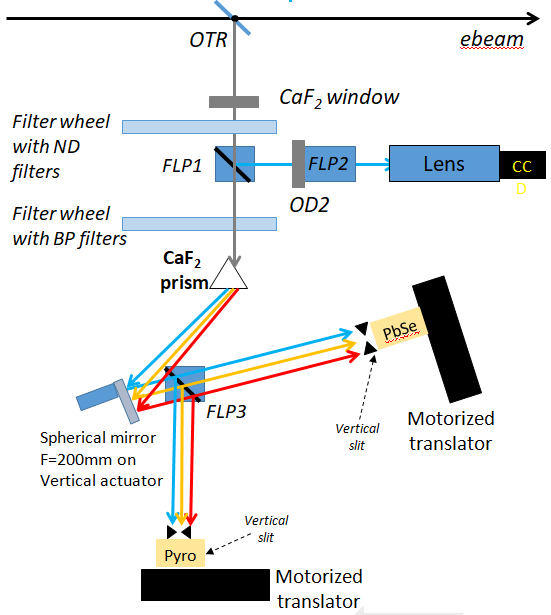}
		\caption{Schematic of the infrared spectrometer. The electron beam passes through an aluminium foil, generating coherent transition radiation (CTR) in the infrared range. This is then dispersed in a CaF\textsubscript{2} prism, and is reflected by a spherical mirror onto either of the two detectors -- a pyrodetector and a PbSe detector. Both of these are placed on motorised translation stages, allowing scanning of the CTR wavelength.} \label{fig:spir}
	\end{center}
\end{figure}

A new type of diagnostic for the microbunching instability, based on the generation of coherent transition radiation (CTR) in the infrared (IR) range, has been developed and commissioned on the FERMI FEL2 line. A $1$\,\si{\micro\metre} aluminium foil is placed in the path of the beam, and the CTR generated by this interaction is then dispersed through a CaF\textsubscript{2} prism and directed onto one of two detectors: a PbSe detector or a pyrodetector. A schematic of the IR spectrometer is shown in Fig.\,\ref{fig:spir}

By moving the detectors on translation stages, the intensity on the detector as a function of CTR wavelength can be measured. These devices were chosen due to their sensitivity ranges, and by using them both, it is possible to cover the wavelength range between $0.25$\,\si{\micro\meter} and $10$\,\si{\micro\meter}, which is most of interest in the context of microbunching in short-wavelength FELs. Indeed, the initial measurements taken using this IR spectrometer were in good agreement with theoretical predictions and previous measurements, giving the strongest microbunching signal around $1$\,\si{\micro\metre}. A brief summary of the initial commissioning results are given in Refs.\,\cite{IPAC2023.MOOG2,FEL2022.WEP24}; a report on a similar device under development at the European XFEL is given in Ref.\,\cite{FEL2019.WEP035}. 

\section{Results}\label{sec:results}

As mentioned above (in Sec.\,\ref{subsec:machine}), two configurations for the transverse optics in the spreader were found for each FEL line. From the injector up to the end of the linac, the machine settings were identical; only two quadrupole magnets in the straight transfer-line section before the spreader dipoles were adjusted in order to increase the value of $\mathcal{H}_{x}$, and the post-spreader quadrupoles were matched to improve the transverse optical beam parameters in the undulator line. 

For each machine setup, the beam transverse Twiss parameters were measured at the end of the linac using the single quad-scan technique \cite{MintyZimmermann}. These values were tracked through the spreader in order to calculate $\mathcal{H}_{x}$ along the line, as shown in Fig.\,\ref{fig:hx}. 

Real-time FEL spectra and pulse intensities were recorded to monitor the performance of the machine under different configurations of the spreader optics. The microbunching instability can cause a reduction in the FEL pulse intensity, and both a broadening of the bandwidth and increase in its jitter, and so these parameters were monitored for each experimental configuration. Herein we describe the bandwidth of the photon pulse as the integrated pixel intensity on the FEL spectrometer containing $76$\,\si{\percent} of the pulse energy \cite{NatPhoton.13.555} (with the spectrometer background subtracted during post-processing). As discussed in Ref.\,\cite{PhysRevAccelBeams.19.050701}, the microbunching instability can create sidebands in the spectrum, disturbing the smooth Gaussian pulse that is characteristic of a seeded FEL. The $76$\,\si{\percent} bandwidth is more sensitive to long tails in the spectrum than the standard full-width half-maximum, and so this parameter can be used as a proxy for the microbunching content in the beam.

\begin{figure*}
	\begin{center}
		\centering
        \subfloat[Pulse intensity]{  
			\includegraphics[width=5.9cm]{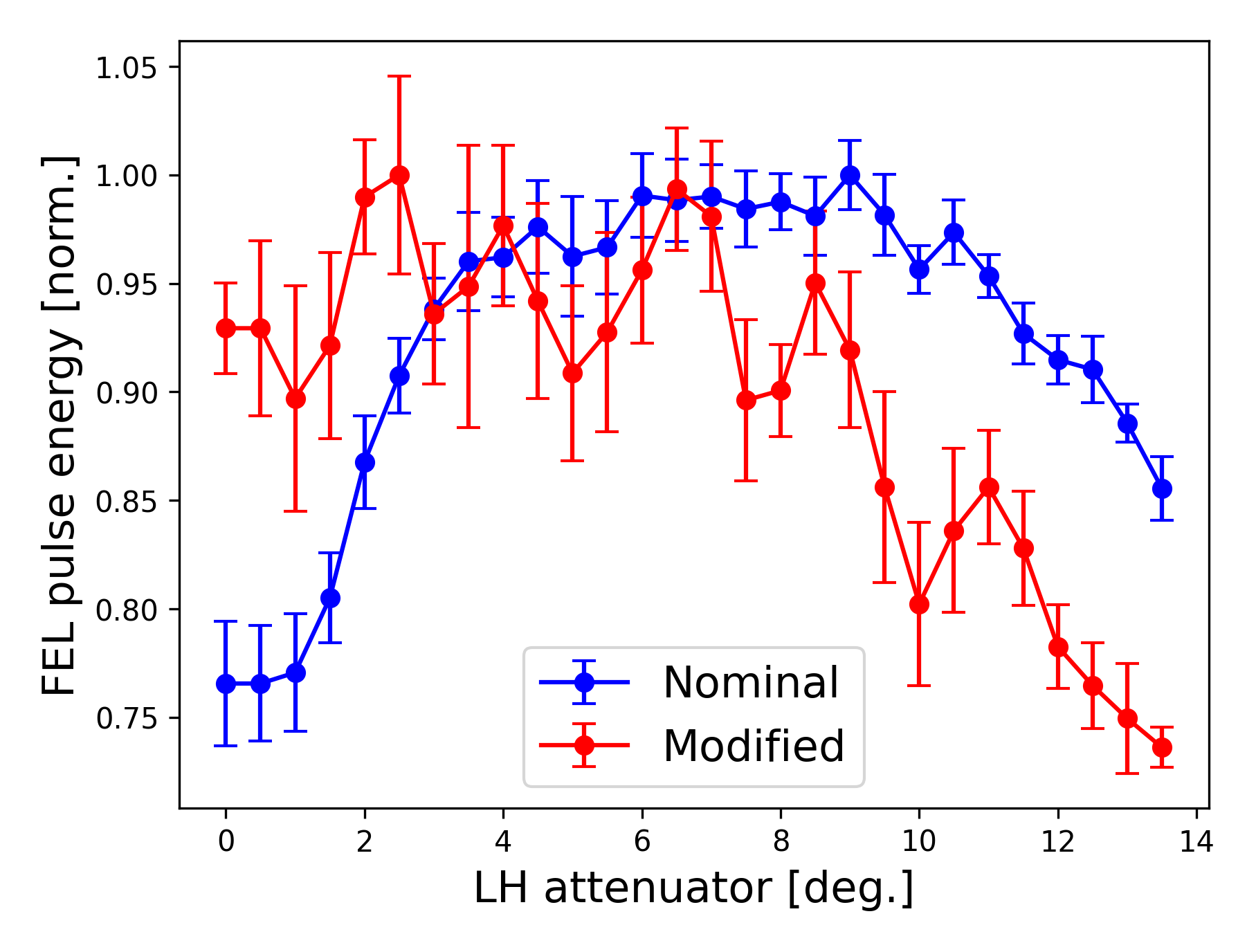} 
			\label{fig:fel1_intensity}}
		\subfloat[Bandwidth]{  
			\includegraphics[width=5.9cm]{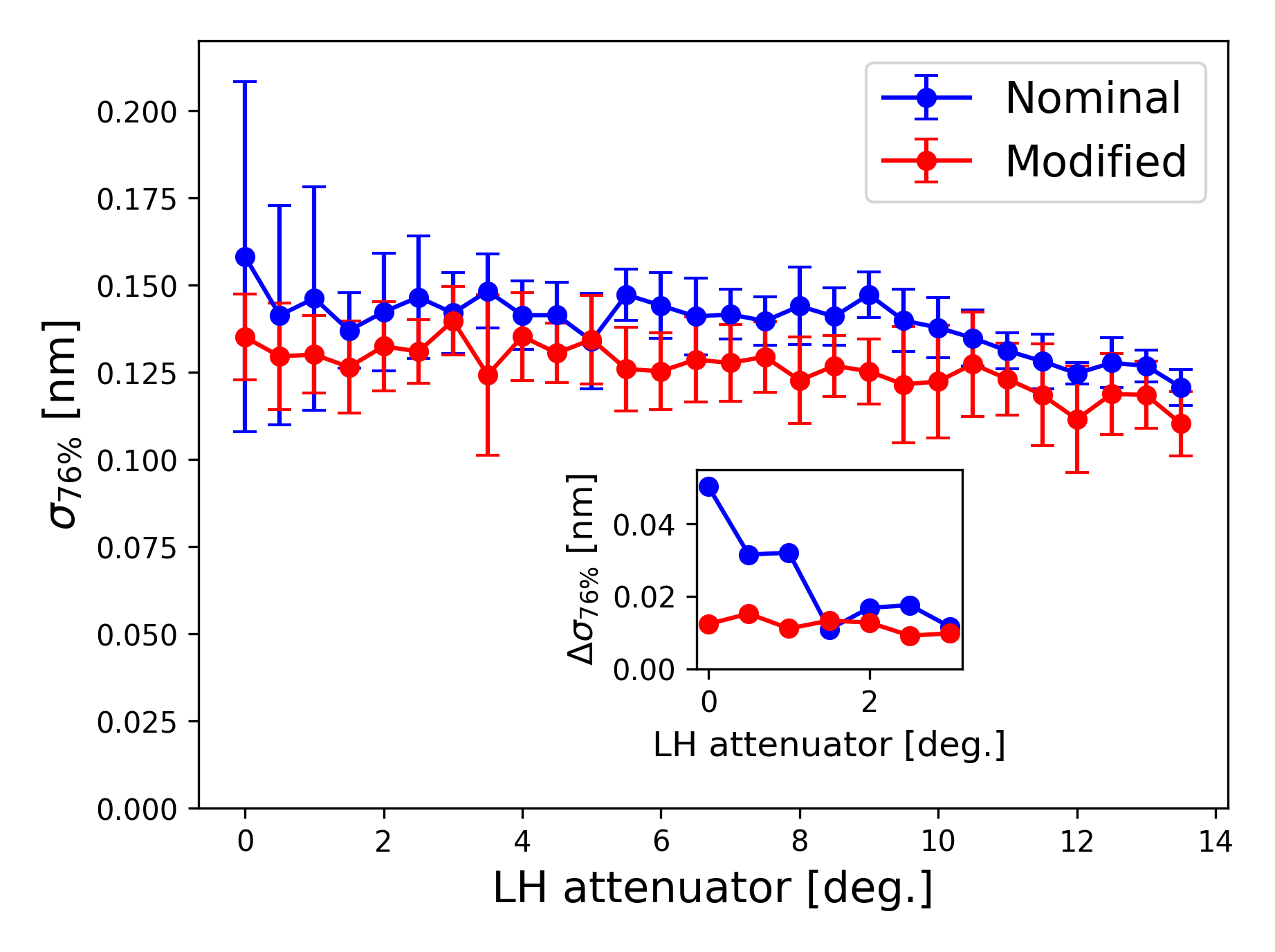} 
			\label{fig:fel1_bandwidth}}
        \subfloat[Spectra]{  
			\includegraphics[width=5.9cm]{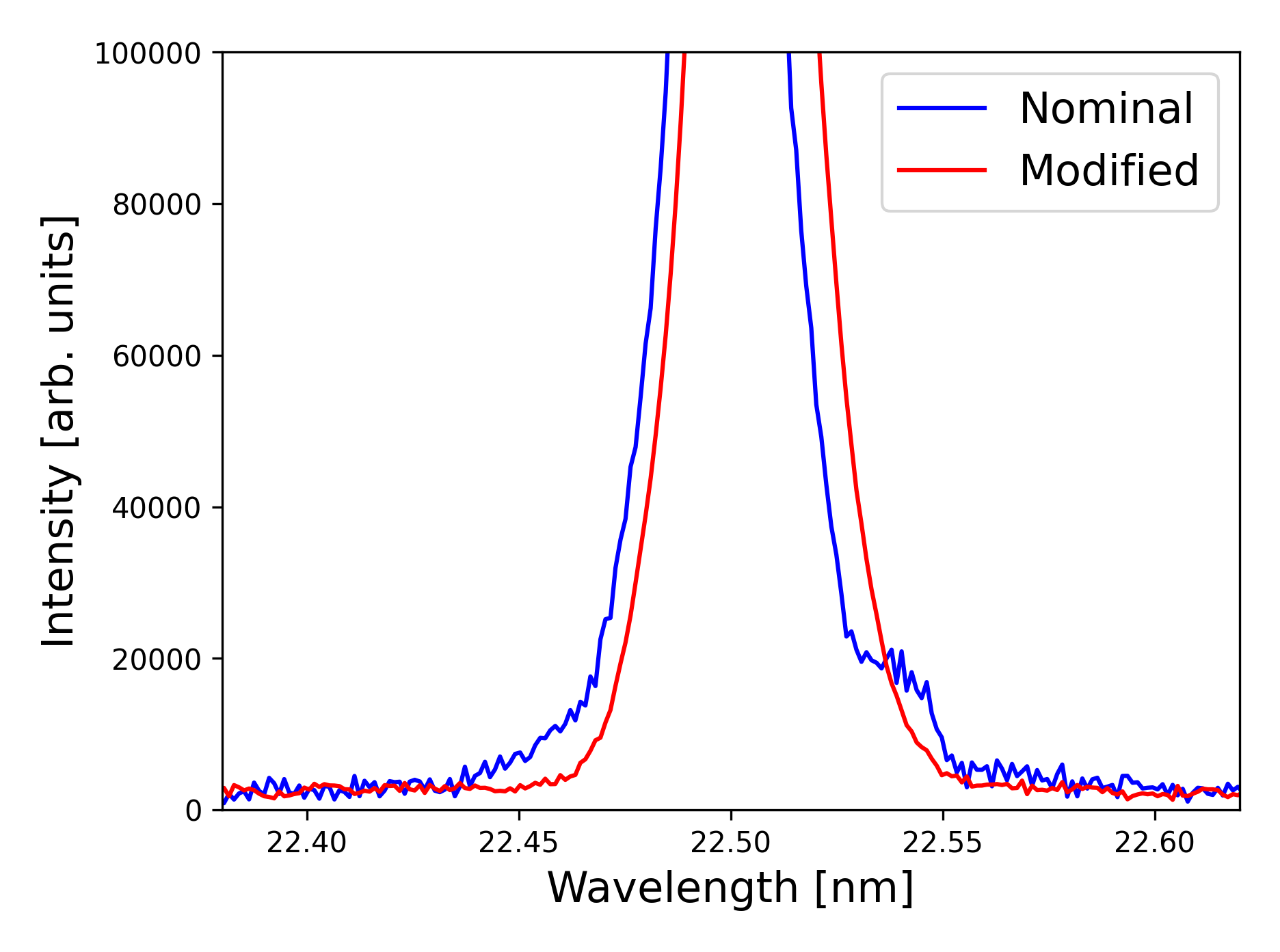} 
			\label{fig:fel1_spectrum}}
		\caption{Performance of FEL1 as a function of laser heater intensity for two different transverse optics configurations in the spreader (left: normalized FEL intensity; middle: $76$\,\si{\percent} bandwidth; right: average spectra for null laser heater action). $20$ shots were measured at each setting of the laser heater attenuator. The radiators were tuned to the 12\textsuperscript{th} harmonic of the seed laser at $270.5$\,\si{\nano\metre}, giving an FEL wavelength of $\approx 22.5$\,\si{\nano\metre}. The inset in Fig.\,\ref{fig:fel1_bandwidth} shows the bandwidth jitter.} \label{fig:fel1_lh_scans}
	\end{center}
\end{figure*}

\begin{figure*}
	\begin{center}
		\centering
        \subfloat[Pulse intensity]{  
			\includegraphics[width=5.9cm]{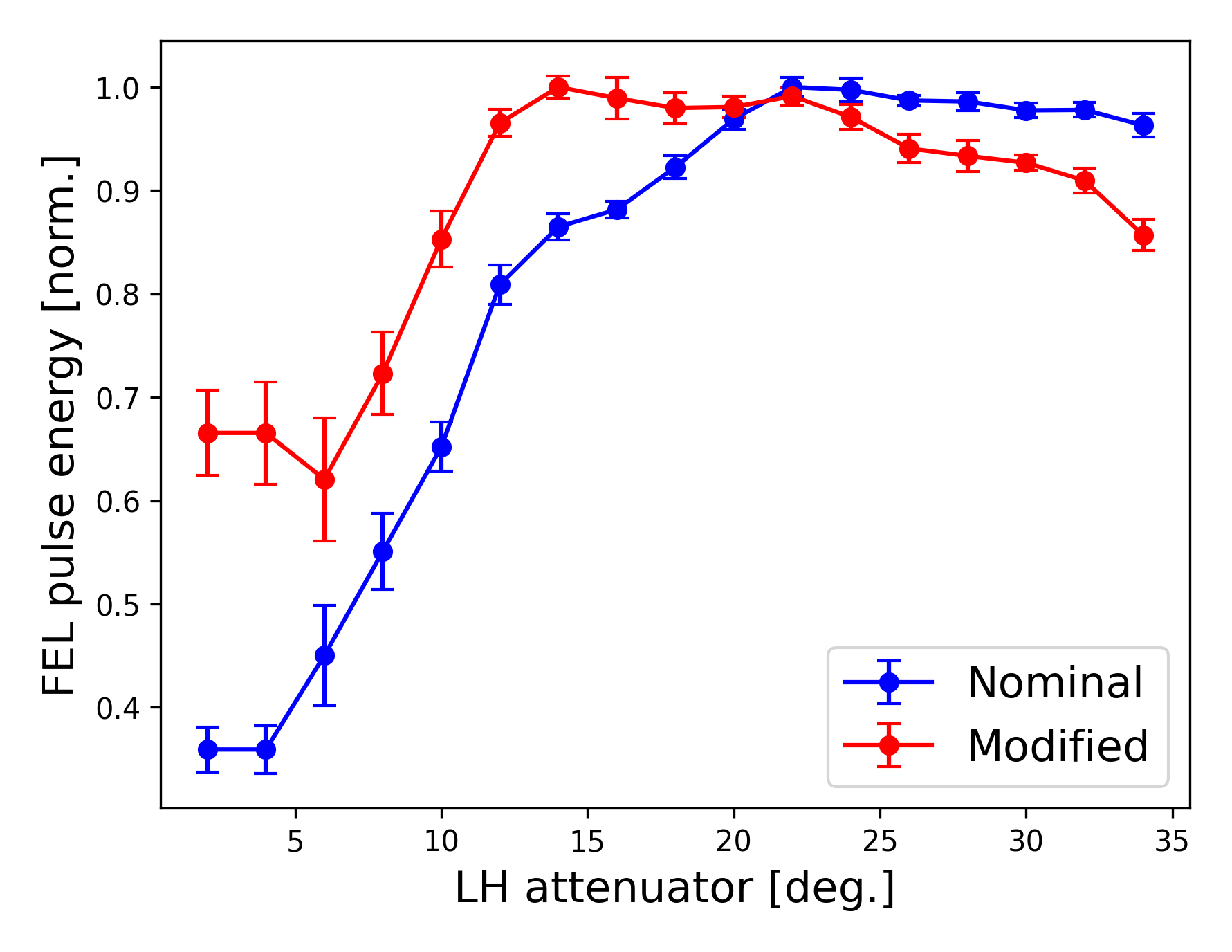} 
			\label{fig:fel2_intensity}}
		\subfloat[Bandwidth]{  
			\includegraphics[width=5.9cm]{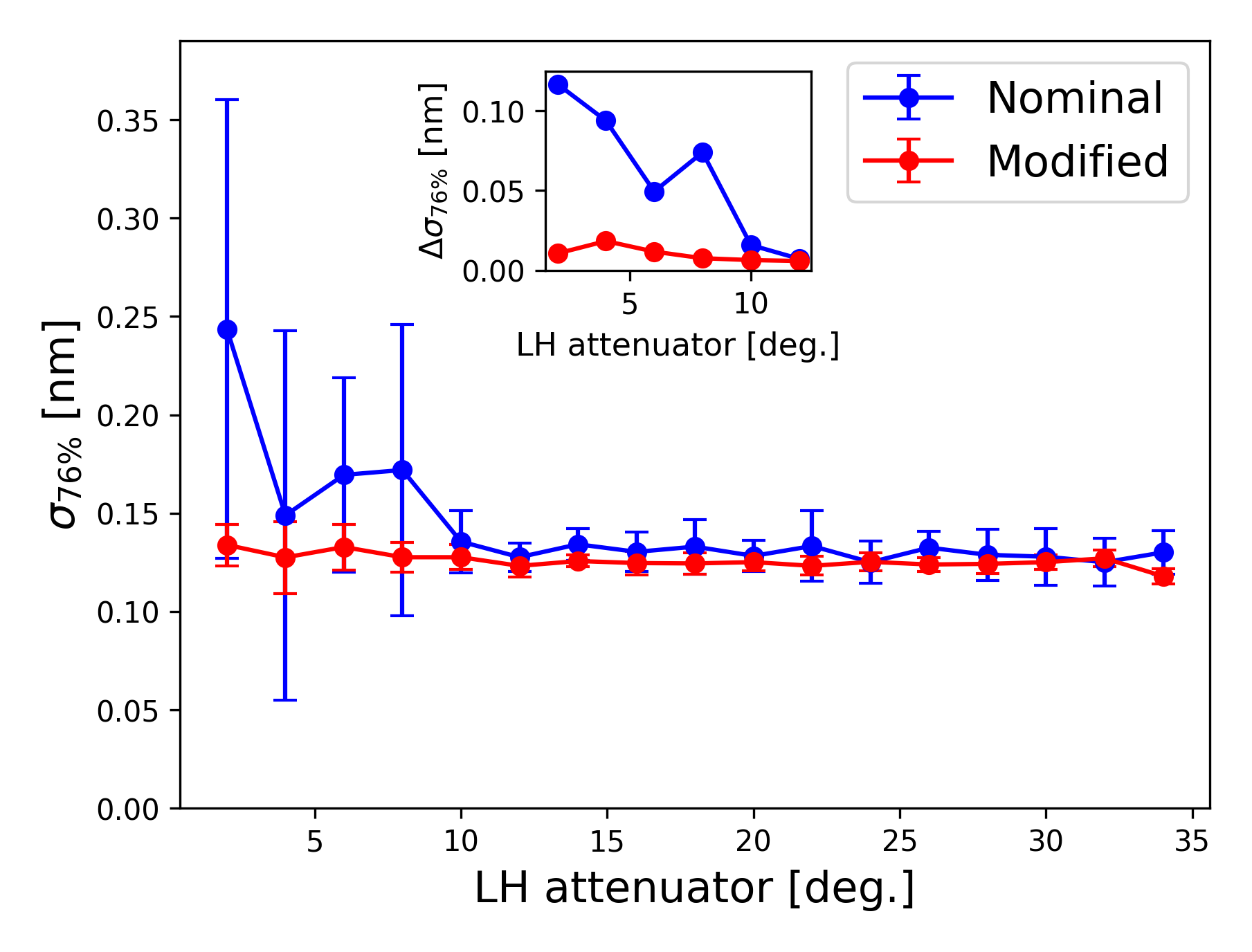} 
			\label{fig:fel2_bandwidth}}
        \subfloat[Spectra]{  
			\includegraphics[width=5.9cm]{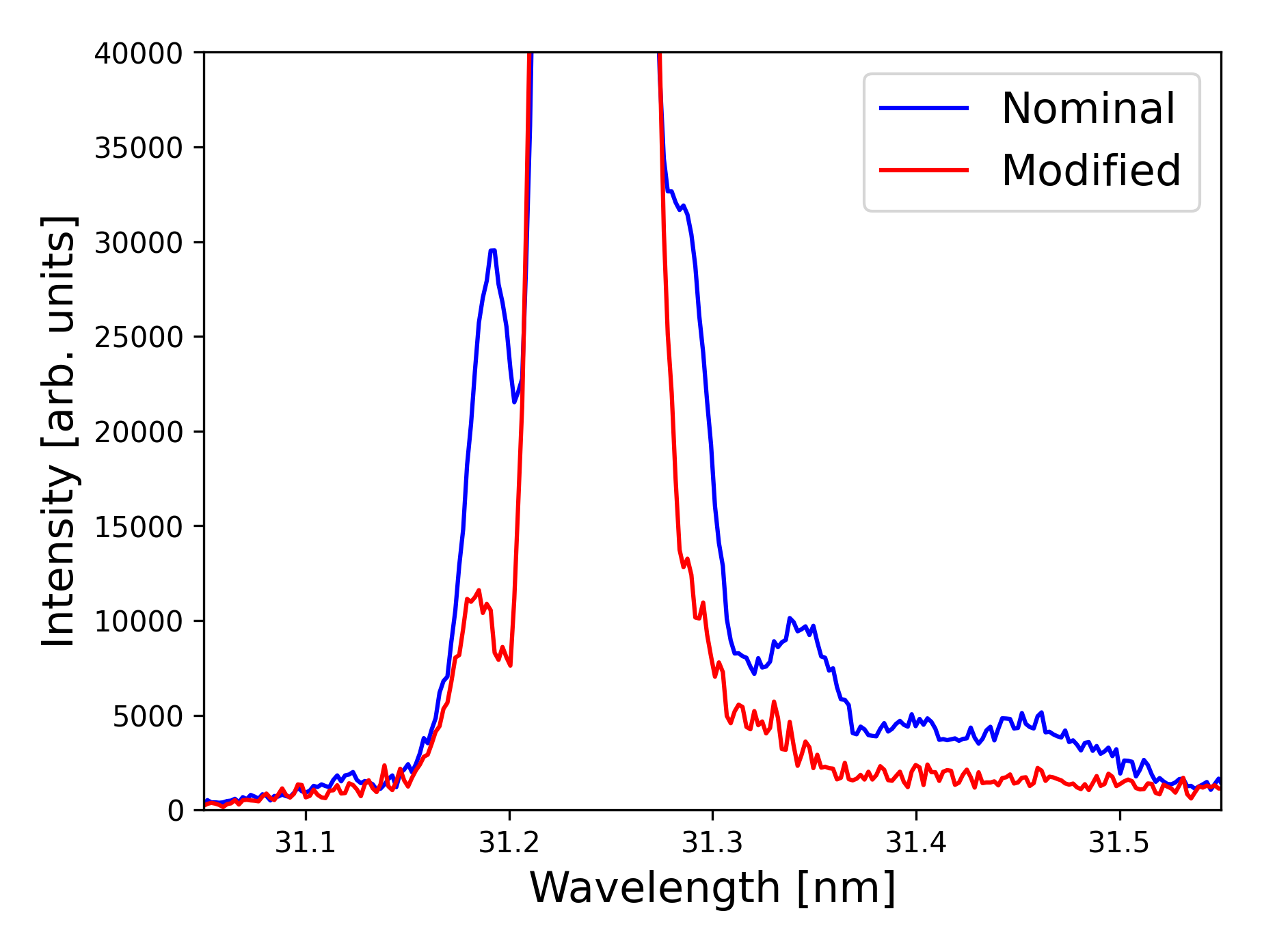} 
			\label{fig:fel2_spectrum}}
		\caption{Performance of FEL2 as a function of laser heater intensity for two different transverse optics configurations in the spreader (left: normalized FEL intensity; middle: $76$\,\si{\percent} bandwidth; right: average spectra for null laser heater action). $20$ shots were measured at each setting of the laser heater attenuator (which in this case had an additional attenuator inserted in order to reduce the pulse energy further). The 1\textsuperscript{st} stage radiators were tuned to the 8\textsuperscript{th} harmonic of the seed laser at $250$\,\si{\nano\metre}, giving an FEL wavelength of $31.25$\,\si{\nano\metre}. The inset in Fig.\,\ref{fig:fel2_bandwidth} shows the bandwidth jitter.}  \label{fig:fel2_lh_scans}
	\end{center}
\end{figure*}


Figures\,\ref{fig:fel1_lh_scans} and \ref{fig:fel2_lh_scans} show measurements of the performance of FEL1 and FEL2, respectively, as a function of the laser heater pulse intensity, which is controlled through the use of a polarizing attenuator (a value of $0$\,\si{\degree} corresponds to null laser heater action). As discussed above, the normalized FEL pulse intensity is shown in Figs.\,\ref{fig:fel1_intensity} and \ref{fig:fel2_intensity}, in order to demonstrate more clearly the effect of microbunching on the beam. On FEL1, the pulse intensity stays relatively close to the maximum value across the scan when the value of $\mathcal{H}_{x}$ is larger in the spreader, gaining only around $5$\,\si{\percent} with a small laser heater action, compared with an increase of $\approx 20$\,\si{\percent} for the nominal case. The analogous case on FEL2 shows a more clear impact of increasing the beam energy spread for both sets of spreader optics, although the increase in FEL pulse energy achieved when introducing the laser heater is around a factor of $2$ larger for the nominal optics, indicating a more significant impact of microbunching on the FEL performance. 

The $76$\,\si{\percent} bandwidths of the FEL pulses across the laser heater scans are shown in Figs.\,\ref{fig:fel1_bandwidth} and \ref{fig:fel2_bandwidth} for FEL1 and FEL2, respectively. On the FEL1 line, the values are comparable for both sets of beam optics; however, the important point to note here is that, for a null laser heater action (with the attenuator at $0$\,\si{\degree}), the variation in the bandwidth is larger when the spreader optics has a smaller value of $\mathcal{H}_{x}$ (as seen in the inset in Fig.\,\ref{fig:fel1_bandwidth}). Given the stochastic nature of the microbunching instability, a larger bandwidth jitter is consistent with a larger presence of microbunching. A similar effect is seen on FEL2 (in the inset of Fig.\,\ref{fig:fel2_bandwidth}), although an additional reduction in the bandwidth by $\approx 40$\,\si{\percent} is also observed for the modified optics. More detailed pictures of the average FEL spectra over $20$ shots for the case of null laser heater action are shown in Figs.\,\ref{fig:fel1_spectrum} and \ref{fig:fel2_spectrum}. On FEL2, sidebands appear more prominently in the FEL spectrum when the nominal spreader optics are used, a feature that has previously been observed and attributed to microbunching \cite{PhysRevAccelBeams.19.050701,PhysRevAccelBeams.22.080702,PhysRevLett.115.214801}; however, the effect of increasing the value of $\mathcal{H}_{x}$ on the average spectrum in FEL1 is small.

\begin{figure}[ht]
	\begin{center}
		\includegraphics[width=86mm]{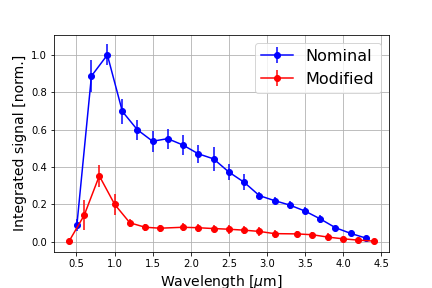}
		\caption{Scan of the integrated signal on the IR spectrometer as a function of wavelength using two different spreader optics configurations. The laser heater pulse energy was set to $100$\,\si{\nano\joule}.} \label{fig:fel2_spir}
	\end{center}
\end{figure}

These two datasets show clearly the response of the FEL to different transverse optics configurations; however, more detailed information about the mircobunching content in the electron beam can be deduced using the IR spectrometer on FEL2. Fig.\,\ref{fig:fel2_spir} shows measurements of the integrated signal on the spectrometer as a function of IR wavelength, controlled through the use of a translation stage. The signal arising from microbunching in the beam is larger by around a factor of $3$ using the nominal beam optics with respect to the modified settings. These measurements also confirm that the modulations in the beam have not been totally removed as a result of modifying the beam transport through the spreader, however; there is still a signal present at around $0.8$\,\si{\micro\metre}, indicating the presence of residual microbunching content in the beam. 

\begin{figure}[ht]
	\begin{center}
		\includegraphics[width=86mm]{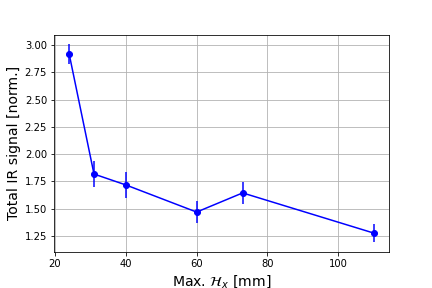}
		\caption{Total signal measured on the IR spectrometer as a function of the maximum value of $\mathcal{H}_{x}$ in the spreader. The laser heater was disabled for these measurements.} \label{fig:spir_hx_mult}
	\end{center}
\end{figure}

A final experiment was conducted in which the limits of this scheme were explored. This was achieved by increasing the value of $\mathcal{H}_{x}$ on the FEL2 line further, under similar operational conditions to those described above (the electron beam energy was $1.37$\,\si{\giga\electronvolt} and the FEL wavelength was $6.38$\,\si{\nano\metre}). As before, the beam emittance was measured before the spreader, and quadrupole magnets before the spreader entrance were used to increase the value of $\mathcal{H}_{x}$. After the spreader, the transverse beam properties were re-matched and the signal on the IR spectrometer was monitored; see Fig.\,\ref{fig:spir_hx_mult}. It can be seen clearly that there is a trend between increasing $\mathcal{H}_{x}$ and reducing the signal on the IR spectrometer, consistent with previous results. However, as $\mathcal{H}_{x}$ was increased, the FEL intensity only increased up to a point, after which the electron beam properties were sufficiently degraded by the modified transport functions such that the FEL performance could not be recovered. This hints at a practical limit of the applicability of this scheme.

\section{Discussion}\label{sec:discussion}

To compare and support the experimental measurements, we use a well-benchmarked semi-analytical Vlasov solver \cite{PhysRevAccelBeams.23.124401,PhysPlasmas.28.013112}, which solves Eq.\,\ref{eq:bunching_factor} and evaluates the microbunching gain for a given beam and lattice. To achieve this, the integration with the kernel function is first split into a discrete sum and can be re-written as a vector-matrix form $b = b_{0} + Kb$. 

The linear integral equation can then be solved by finding the inverse of the matrix $(\boldsymbol{I} - K)$, with $\boldsymbol{I}$ the identity matrix. The number of numerical meshes must be sufficient to ensure that the results are converged in the presence of collective effects. The capabilities of the Vlasov solver include a list of relevant collective effects including the impedances due to CSR (both steady-state and transient \cite{NIMA.398.2-3.373}) and LSC, and intrabeam scattering models. 

This Vlasov solver is applicable for constant beam energy and with beam acceleration. The $6 \times 6$ linear transport matrix along the beamline is adopted from the \textsc{Elegant} tracking code \cite{APS-LS-287}. The solver provides a fast, efficient, and satisfactory estimate for microbunching dynamics for a linear beamline lattice. 

\begin{figure}[ht]
	\begin{center}
		\centering
		\subfloat[FEL1]{  
			\includegraphics[width=8.6cm]{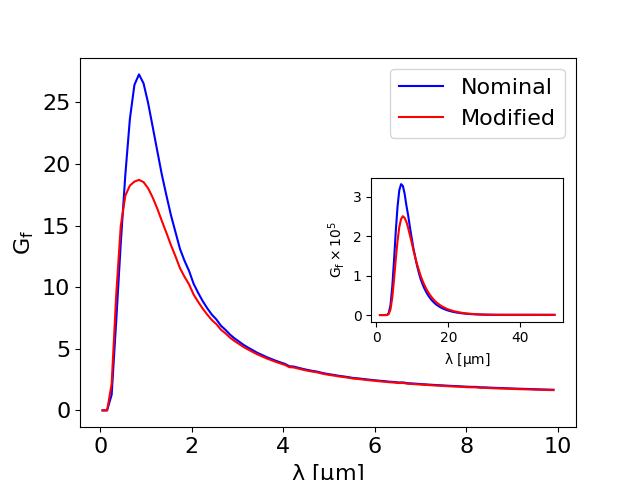} 
			\label{fig:fel1_mbi_gain}}
            \vfill
		\subfloat[FEL2]{  
			\includegraphics[width=8.6cm]{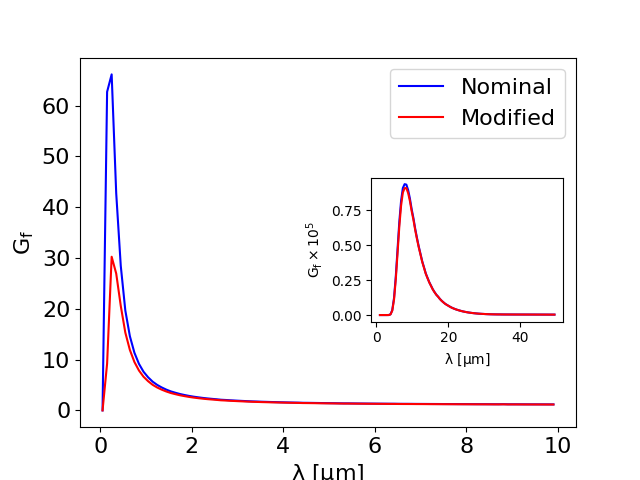} 
			\label{fig:fel2_mbi_gain}}
		\caption{Calculated microbunching gain (due to CSR and LSC) as a function of compressed modulation wavelength along the spreader for the nominal (blue) and the modified optics (red). The inset figures show the final microbunching gain along the full line. The predicted peak modulation wavelengths, from the entire line and the spreader only, are close to those from experimental measurements.} \label{fig:mbi_calculated}
	\end{center}
\end{figure}

For the two pairs of datasets on each FEL line, the development of the microbunching gain along the beamline was calculated using Eq.\,\ref{eq:bunching_factor}. Each lattice configuration was simulated using \textsc{Elegant}, starting at the laser heater, with values for the initial beam properties at the exit of the injector provided by a General Particle Tracer simulation \cite{GPT}. The most pertinent property for the initial distribution was a slice energy spread of $2$\,\si{\kilo\electronvolt}, which is consistent with previous work \cite{NewJPhys.22.083053}. Given the crucial nature of the transverse beam properties in the spreader, the simulated beam was re-matched at the spreader entrance based on measured values of the Twiss properties at this location. By applying the semi-analytical treatment of the microbunching gain along each lattice element, based on the simulated electron beam properties, the microbunching gain at the exit of the linac and along the entire line (including the spreader) was calculated. The results are shown in Fig.\,\ref{fig:mbi_calculated}. The calculation of the gain in the spreader was started at the emittance measurement station at the exit of the linac, in order to show more clearly the effect of the modification to the beam optics. 

In solving Eq.\,\ref{eq:bunching_factor} the mesh number for integration along $s$ is set to ensure sufficient numerical convergence. In the semi-analytical calculation both the LSC and CSR impedances (both steady-state and transient \cite{NIMA.398.2-3.373}) are included. For calculations along the entire line, the mesh number is set to $6000$, the initial beam parameters are given in Table\,\ref{table:machine_setup}, and the lattice $6 \times 6$ transport functions from \textsc{Elegant} are used. For the spreader-only calculations, the mesh number is set to $2000$ and the initial beam parameters are taken from the experimental measurements, re-matched at the linac exit. Specifically, the slice beam energy spread is set at $70$\,\si{\kilo\electronvolt} and the beam emittance remains the same as the initial value. Here we remark that, in the semi-analytical calculation, the unperturbed beam dynamics are assumed based on linear transport; the quadratic chirp and/or phase space distortion, accumulated along the beamline, is overlooked, which may lead to an overestimation of the resultant microbunching gain. However, the relative trend should be retained between the nominal and modified lattices.

The transverse Landau damping term $LD_{\perp}$ (Eq.\,\ref{eq:landau_damping_transverse}) can be correlated with the improvement in FEL performance. The relevant experimental parameters are summarized in Table\,\ref{table:parameters}. The longitudinal Landau damping term $LD_{\parallel}$ (given by the first exponential term of Eq.\,\ref{eq:kernel}, although fixed for both sets of spreader optics, is also shown. 

\begin{table}[bth!]
\centering
\caption{Beam and spreader parameters for the experiments on each FEL line. The larger values of $\mathcal{H}_{x}$ correspond to the datasets labelled \lq Modified Optics\rq.} 
\label{table:parameters}
\begin{tabular}{llll}
    \hline\hline
    \textbf{Parameter} & \textbf{FEL1} & \textbf{FEL2}  & \textbf{Unit} \\
    \hline\hline
    Beam energy & 1.238 & 1.533 & \si{\giga\electronvolt} \\
    $\mathcal{H}_{x,\text{max}}$ & $16\rightarrow 30$ & $25\rightarrow 40$ &  \si{\milli\metre} \\
    $R_{56}$ & 1.24 & 0.78 & \si{\milli\metre} \\
    $LD_{\perp}$ & $0.85\rightarrow 0.73$ & $0.84\rightarrow 0.76$ & \hbox{--}\\
    $LD_{\parallel}$ & 0.50 & 0.84 & \hbox{--} \\
    \hline
\end{tabular}
\end{table}  

It can be seen that there is a correspondence between the increased value of $\mathcal{H}_{x}$, and therefore enhanced transverse Landau damping, and the reduction in the microbunching gain along the spreader. The scans of the laser heater pulse energy, demonstrating an increase in brightness and a reduced sideband intensity in the FEL spectrum for low values of added energy spread with a larger value of $\mathcal{H}_{x}$, show that the experimental measurements are in agreement with arguments from theory and simulation. 

Although the reduction in the microbunching gain (Fig.\,\ref{fig:mbi_calculated}) is moderate, it demonstrates a possible trend towards the removal of microbunching-driven modulations in the beam longitudinal phase space cumulated along energy-dispersive regions, therefore impacting the total gain. As there are indications of residual microbunching structure in the beam, evidenced in particular in the measurements taken using the IR spectrometer (Figs.\,\ref{fig:fel2_spir} and \ref{fig:spir_hx_mult}), and also in the measurements of the FEL spectra (Fig.\,\ref{fig:fel2_spectrum}), there is potential scope for further improvements in the optics-based damping of microbunching gain. This could be achieved either by increasing the transverse Landau damping in other dispersive regions, or through coupling this scheme with the previously demonstrated method of amplifying the longitudinal Landau damping term by changing the $R_{56}$ in the spreader \cite{PhysRevAccelBeams.23.110703}. Recent experiments performed at FERMI which involved increasing both $\mathcal{H}_{x}$ and $R_{56}$ in the spreader line demonstrated a further reduction in the microbunching content in the beam, leading to an improvement of the FEL performance with a lower setting of the laser heater.

\section{Conclusion}\label{sec:conclusion}

The microbunching instability has long been recognized as a considerable barrier to achieving longitudinal coherence in FELs. Although laser heaters have proven to be very effective in suppressing unwanted modulations in the longitudinal phase space of the electron beam, alternatives to this method are promising avenues of research, as they can provide reversible schemes for mitigating the instability. In this paper, we have demonstrated improved performance of the FERMI FEL by modifying the transverse beam optics upon entering a dispersive region. While the transport matrix for the spreader remained fixed, the initial conditions -- namely, $\beta_{x}$ and  $\alpha_{x}$ -- at the spreader entrance were changed through the use of quadrupole magnets. In doing so, the value of $\mathcal{H}_{x}$ along the spreader was increased, thereby providing an appreciable amount of damping of the microbunching gain.

Experimentally, we observed an improvement in the FEL performance for a larger value of $\mathcal{H}_{x}$ in the spreader; in particular, the contribution of sidebands from the central FEL wavelength to the total pulse intensity was reduced, evidenced through a reduction in the bandwidth. Additionally, the bandwidth jitter was reduced for low laser heater pulse intensity. The intensity of the IR spectrum produced by the beam upon passing through a screen in the FEL2 line was also diminished for the modified optics, which is consistent with the FEL measurements. These results were also corroborated with a semi-analytic evaluation of the evolution of the microbunching gain along the linac and spreader lines, demonstrating a damping of the gain which agrees well with the measurements.

A limitation of this scheme is that the value of $\mathcal{H}_{x}$ cannot be made arbitrarily large; while the possibility of entirely removing modulations in the bunch using this method could be possible in principle, the performance of the FEL must also be taken into account. If the beam transport is not kept under reasonable control through the spreader, it would not be possible to match the beam into the FEL, leading either to unacceptable beam losses or a degradation in the beam properties. Indeed, the measurements of the IR spectrometer signal demonstrated that some residual microbunching could have persisted in the beam in the cases where the value of $\mathcal{H}_{x}$ was large. 

However, these results represent an advance in schemes for damping the microbunching instability that rely on hardware that is readily available in any accelerator, and demonstrate the possibility of more advanced spreader designs and beam optics configurations that can take advantage of this effect. This technique, merged with longitudinal phase mixing, can be complementary to the laser heater by reducing this tool to low pulse energies, which is of crucial importance for \si{\mega\hertz}-rate superconducting FELs targeting full coherence in the x-ray range. Given the increasingly stringent requirements on the electron beam energy spread demanded by advanced FEL schemes, the method described in this paper demonstrates an alternative route towards improving the performance of these machines.

\section*{Acknowledgment}

One of the authors (C.-Y. T.) is supported by the Fundamental Research Funds for the Central Universities (HUST) under Project No. 2021GCRC006 and National Natural Science Foundation of China under project No. 12275094.

\bibliographystyle{unsrt}
\bibliography{references}

\end{document}